\documentclass[onecolumn,preprintnumbers,amsmath,amssymb,notitlepage]{revtex4}
\usepackage{graphicx}
\usepackage{epsfig}
\usepackage{amsmath}
\usepackage{amssymb}
\usepackage{dcolumn}

\begin{document}

\title{Conformal Gravity on Noncommutative Spacetime}

\author{Martin Kober${}^1$}
\email{kober@fias.uni-frankfurt.de}

\affiliation{Frankfurt Institute for Advanced Studies (FIAS),
Johann Wolfgang Goethe-Universit\"at,
Ruth-Moufang-Strasse 1, 60438 Frankfurt am Main, Germany}

\date{\today}

\begin{abstract}
Conformal gravity on noncommutative spacetime is considered in this paper. The presupposed gravity action consists
of the Brans-Dicke gravity action with a special prefactor of the term, where the Ricci scalar couples to the
scalar field, to maintain local conformal invariance and the Weyl gravity action. The commutation relations
between the coordinates defining the noncommutative geometry are assumed to be of canonical shape. Based on the
moyal star product, products of fields depending on the noncommutative coordinates are replaced by generalized
expressions containing the usual fields and depending on the noncommutativity parameter. To maintain invariance
under local conformal transformations with the gauge parameter depending on noncommutative coordinates, the fields
have to be mapped to generalized fields by using Seiberg-Witten maps. According to the moyal star product and the
thus induced Seiberg-Witten maps the generalized conformal gravity action is formulated and the corresponding
field equations are derived.
\end{abstract}

\maketitle

\section{Introduction}

The idea of noncommutative geometry was first considered in \cite{Snyder:1946qz}. Since that time noncommutative geometry
has become one of the most important concepts in fundamental physics. There exist two approaches to formulate
field theories on a spacetime with noncommuting coordinates. One possibility is the application of the the moyal star
product, which was developed in \cite{Groenewold:1946kp} and \cite{Moyal:1949sk}. It is based on Weyl quantization
and enables to convert products of fields depending on noncommutative coordinates to generalized products of these fields depending on usual coordinates. These generalized products of course contain the noncommutativity parameter defining the noncommutativity algebra. Another way to treat noncommutative field theories is the coherent state approach, developed in \cite{Smailagic:2003rp},\cite{Smailagic:2003yb},\cite{Smailagic:2004yy} and extended by incorporating noncommuting momenta
in \cite{Kober:2010um}. In the coherent state approach are defined creation and annihilation operators from the noncommutative
coordinates with respect to which coherent states are defined. By using these coherent states, plane waves depending on
noncommutative coordinates can be expressed as generalized plane waves depending on usual coordinates.
However, in this paper conformal gravity on noncommutative spacetime is considered and the moyal star product approach is
taken as a basis. Especially, a gravity action is presupposed, which consists of a special Brans-Dicke gravity action,
where the prefactor of the coupling term between the Ricci scalar and the scalar field is chosen in such a way to maintain
local conformal invariance, on the one hand and of the Weyl gravity action on the other hand. Brans-Dicke theory was first
developed in \cite{Brans:1961sx},\cite{Brans:1962zz}. Conformally invariant theories, where the gravitational constant arises
from the scalar field, were considered in \cite{Fujii:1982ms},\cite{Foot:2007iy}. In \cite{Maldacena:2011mk} has recently
been considered the relation between Einstein gravity and conformal gravity.

A precondition to maintain invariance under a certain local symmetry group on noncommutative spacetime, symmetry under
conformal transformations in the special case of this paper, is the introduction of Seiberg-Witten maps, which was devised
in \cite{Seiberg:1999vs}. Seiberg-Witten maps relate the usual fields in the corresponding theory to generalized fields,
which have to be used instead of the usual fields. The reason is that according to the moyal star product, on noncommutative spacetime additional terms appear for products of fields and this holds also for the products in the transformation rules of
the fields, since the transformation parameters also have to be assumed to depend on the noncommutative coordinates, if a spacetime with noncommutative coordinates is presupposed.
Based on these concepts, the moyal star product as well as the Seiberg-Witten maps, in \cite{Calmet:2004ii},\cite{Calmet:2005qm},\cite{Calmet:2006iz},\cite{Calmet:2005mc} were considered spacetime
symmetries and general relativity on noncommutative spacetime. Other considerations concerning gravity on noncommutative spacetime can be found in \cite{Chamseddine:1992yx},\cite{Madore:1993br},\cite{Heller:1999zz},\cite{Heller:2003gz},\cite{Heller:2005qa},\cite{Moffat:2000gr},\cite{Moffat:2000fv},\cite{Okawa:2000sh},\cite{Avramidi:2003mk},\cite{Yang:2004vd},\cite{Aschieri:2005yw},\cite{Harikumar:2006xf},\cite{Banerjee:2007th},\cite{MullerHoissen:2007xy},\cite{Vassilevich:2009cb},\cite{Faraoni:1998qx},\cite{Aschieri:2009ky},\cite{Miao:2010kr},\cite{Nicolini:2010nb},\cite{Miao:2010yq} for example and a noncommutative tetrad
field was considered in \cite{Kober:2011am}.
In this paper, the idea of formulating a theory on noncommutative spacetime, which contains a certain symmetry,
by introducing Seiberg-Witten maps is transferred to conformally invariant gravity.

The paper is structured as follows: At the beginning, the conformally invariant gravity theory and the transition to
noncommuting coordinates are presented. This means that the products within the action and within the transformation
rules of the fields become products between fields depending on noncommuting coordinates. These products are converted
to moyal star products. To maintain gauge invariance, the fields have to be replaced by generalized fields, which can
be expressed in terms of the usual fields by using Seiberg-Witten maps. Accordingly the Seiberg-Witten maps corresponding
to conformal transformations of the tetrad field and the scalar field are constructed and it is shown that they fulfil
the general condition defining Seiberg-Witten maps. After this the corresponding generalized geometrical quantities,
the determinant of the tetrad field, the connection referring to the gravitational field and the Riemann tensor on
noncommutative spacetime, are calculated to obtain the conformally invariant combined Brans-Dicke and Weyl gravity
action on noncommutative spacetime under presupposition of the moyal star product approach. Finally, the corresponding
field equations are derived by varying the obtained generalized gravity action with respect to the tetrad field and
the scalar field.

\section{Conformal Gravity Action}

Before considering noncommutative spacetime, the presupposed gravity action is presented and it is reformulated
in terms of the tetrad field, since this will simplify the treatment of the noncommutative case. In this paper
is presupposed a gravity action consisting of a special Brans-Dicke gravity action and the Weyl gravity action,
which looks as following, if it is formulated on usual spacetime: 

\begin{equation}
S_C=\int d^4 x \sqrt{-g}\left(\frac{1}{2}g^{\mu\nu}\partial_\mu \phi \partial_\nu \phi+\frac{1}{12} R \phi^2
+C_{\mu\nu\rho\sigma}C^{\mu\nu\rho\sigma}\right),
\label{usual_conformal_gravity}
\end{equation}
where $\phi$ denotes a scalar field, $g^{\mu\nu}$ denotes the metric, $g$ its determinant $\det g_{\mu\nu}$, $R$ the
Ricci scalar $R=g^{\mu\nu}R_{\mu\nu}=g^{\mu\nu}R_{\mu\rho\nu}^{\ \ \ \ \rho}$, where $R_{\mu\nu}$ denotes the Ricci tensor
and $R_{\mu\nu\rho\sigma}$ the Riemann tensor, and $C_{\mu\nu\rho\sigma}$ denotes the conformal Weyl tensor depending
on the metric tensor and the quantities constructed from it according to

\begin{eqnarray}
C_{\mu\nu\rho\sigma}&=&R_{\mu\nu\rho\sigma}-\frac{2}{D-2}\left(g_{\mu\rho}R_{\sigma\nu}-g_{\mu\sigma}R_{\rho\nu}
-g_{\nu\rho}R_{\sigma\mu}+g_{\nu\sigma}R_{\rho\mu}\right)+\frac{2}{\left(D-1\right)\left(D-2\right)}
\left(R g_{\mu\rho}g_{\sigma\nu}-R g_{\mu\sigma}g_{\rho\nu}\right)\nonumber\\
&=&R_{\mu\nu\rho\sigma}-\left(g_{\mu\rho}R_{\sigma\nu}-g_{\mu\sigma}R_{\rho\nu}
-g_{\nu\rho}R_{\sigma\mu}+g_{\nu\sigma}R_{\rho\mu}\right)
+\frac{1}{3}\left(R g_{\mu\rho}g_{\sigma\nu}-R g_{\mu\sigma}g_{\rho\nu}\right),
\label{conformal_tensor}
\end{eqnarray}
if the spacetime dimension $D$ is assumed to be $D=4$.
The gravity action ($\ref{usual_conformal_gravity}$) is invariant under the following local
conformal transformations:

\begin{equation}
g_{\mu\nu} \rightarrow e^{2\lambda(x)} g_{\mu\nu},\quad \phi \rightarrow e^{-\lambda(x)}\phi,
\label{conformal_transformations}
\end{equation}
implying the following transformations for the Ricci scalar and the squared Weyl tensor:

\begin{eqnarray}
R &\rightarrow& e^{-2\lambda(x)}\left[R-2\left(D-1\right)\frac{1}{\sqrt{-g}}
\partial_\mu\left(\sqrt{-g}g^{\mu\nu}\partial_\nu \lambda(x)\right)
-\left(D-1\right)\left(D-2\right)g^{\mu\nu}\partial_\mu \lambda(x) \partial_\nu \lambda(x)\right]\nonumber\\
&&=e^{-2\lambda(x)}\left\{R-6\left[\frac{1}{\sqrt{-g}}
\partial_\mu\left(\sqrt{-g}g^{\mu\nu}\partial_\nu \lambda(x)\right)
+g^{\mu\nu}\partial_\mu \lambda(x)\partial_\nu \lambda(x)\right]\right\},\ {\rm since}\ D=4\nonumber\\
C_{\mu\nu\rho\sigma}C^{\mu\nu\rho\sigma} &\rightarrow& e^{-4\lambda(x)} C_{\mu\nu\rho\sigma}C^{\mu\nu\rho\sigma},
\label{transformation_Ricci_Weyl}
\end{eqnarray}
which are relevant for the conformal gravity action ($\ref{usual_conformal_gravity}$). That $g_{\mu\nu}$
is transformed to $e^{2\lambda(x)} g_{\mu\nu}$ implies of course that the inverse metric $g^{\mu\nu}$ is
transformed to $e^{-2\lambda(x)} g^{\mu\nu}$ and that the square root of the negative determinant of the
metric $\sqrt{-g}$ is transformed to $e^{4\lambda(x)}\sqrt{-g}$. The prefactor of the coupling term between the
Ricci scalar and the scalar field in ($\ref{usual_conformal_gravity}$) is chosen in such a way that the
transformation of this term implied by ($\ref{transformation_Ricci_Weyl}$) cancels the transformation
of the kinetic term of the scalar field $\phi$. The Weyl gravity action is invariant under conformal transformations,
because the appearing factor after the conformal transformation of the squared Weyl tensor ($\ref{transformation_Ricci_Weyl}$),
from which it is constructed, exactly cancels the factor of the square root of the negative determinant
of the metric. With respect to the further considerations, it will be useful to rewrite the conformal
gravity action ($\ref{usual_conformal_gravity}$) and reexpress it by using the tetrad formulation of
the gravitational field. The tetrad field, denoted by $E_\mu^m$ in this paper, where the Greek letter
denotes the spacetime index and the Latin letter denotes the Lorentz index, is related to the metric
as usual according to

\begin{equation}
g_{\mu\nu}=E_\mu^m E_\nu^n \eta_{mn},
\label{tetrad_field}
\end{equation}
\newpage\noindent
where $\eta_{mn}$ denotes the Minkowski metric of flat spacetime. This means that the conformally invariant
gravity action ($\ref{usual_conformal_gravity}$) expressed by the tetrad field ($\ref{tetrad_field}$) reads

\begin{equation}
S_C=\int d^4 x\ E\left[\frac{1}{2}E^{\mu m}E^\nu_m \partial_\mu \phi \partial_\nu \phi+\frac{1}{12} R\left(E\right)\phi^2
+C_{\mu\nu\rho\sigma}\left(E\right)C^{\mu\nu\rho\sigma}\left(E\right)\right],
\label{usual_conformal_gravity_tetrad}
\end{equation}
where $E= \det E_\mu^m$ and the Riemann tensor $R_{\mu\nu}^{ab}$ is now defined by the spin connection
$\omega_\mu^{ab}$ according to

\begin{equation}
R_{\mu\nu}^{ab}(E)=\partial_\mu \omega_\nu^{ab}(E)-\partial_\nu \omega_\mu^{ab}(E)
+\omega_\mu^{ac}(E)\omega_\nu^{cb}(E)-\omega_\nu^{ac}(E)\omega_\mu^{cb}(E),
\end{equation}
the spin connection depends on the tetrad field in the following way:

\begin{equation}
\omega_\mu^{ab}(E)=2 E^{\nu a} \partial_\mu E_\nu^b-2 E^{\nu b} \partial_\mu E_\nu^a
-2 E^{\nu a} \partial_\nu E_\mu^b+2 E^{\nu b} \partial_\nu E_\mu^a
+E_{\mu c} E^{\nu a} E^{\sigma b} \partial_\sigma E_\nu^c
-E_{\mu c} E^{\nu a} E^{\sigma b} \partial_\nu E_\sigma^c,
\end{equation}
and the Ricci tensor and the Ricci scalar are accordingly defined as: $R_\mu^a=E^\nu_b R_{\mu\nu}^{ab}$ and
$R=E^\mu_a E^\nu_b R_{\mu\nu}^{ab}$. The conformal tensor ($\ref{conformal_tensor}$) expressed in terms of
the tetrad field reads

\begin{eqnarray}
C_{\mu\nu\rho\sigma}(E)&=&E_{\nu a} E_{\sigma b} R_{\mu\rho}^{ab}(E)
-\left[E_\mu^a E_{\rho a} E_{\nu b} E^\tau_c R_{\sigma\tau}^{bc}(E)
-E_\mu^a E_{\sigma a} E_{\nu b} E^\tau_c R_{\rho\tau}^{bc}(E)
-E_\nu^a E_{\rho a} E_{\mu b} E^\tau_c R_{\sigma\tau}^{bc}(E)
\right.\nonumber\\&&\left.
+E_\nu^a E_{\sigma a} E_{\mu b} E^\tau_c R_{\rho\tau}^{bc}(E)\right]
+\frac{1}{3}\left[E^\tau_b E^\upsilon_c R_{\tau\upsilon}^{bc}(E) E_\mu^a E_{\rho a} E_\sigma^a E_{\nu a}
-E^\tau_b E^\upsilon_c R_{\tau\upsilon}^{bc}(E) E_\mu^a E_{\sigma a} E_\rho^a E_{\nu a}\right].
\end{eqnarray}
The conformal gravity action expressed by the tetrad field ($\ref{usual_conformal_gravity_tetrad}$)
is invariant under the conformal transformations corresponding to ($\ref{conformal_transformations}$),
which are of the following form:

\begin{equation}
E_\mu^m \rightarrow e^{\lambda(x)} E_\mu^m=E_\mu^m+\lambda(x) E_\mu^m+\mathcal{O}\left(\lambda^2\right),\quad
\phi \rightarrow e^{-\lambda(x)}\phi=\phi-\lambda(x)\phi+\mathcal{O}\left(\lambda^2\right),
\label{conformal_transformations_tetrad}
\end{equation}
implying that the inverse tetrad $E^\mu_m$ is transformed to $e^{-\lambda(x)}E^\mu_m$.

\section{Transition to Noncommutative Spacetime and Seiberg-Witten Maps}

In this section the conformal gravity action presented in the last section will be formulated on noncommutative spacetime.
This means that a transition from commuting to noncommuting coordinates has to be performed, $x^\mu \rightarrow \hat x^\mu$.
The noncommutative geometry represented by the noncommuting coordinates $\hat x^\mu$ is assumed to be described by
canonical commutation relations between the coordinates, which are of the following shape:

\begin{equation}
\left[\hat x^\mu,\hat x^\nu \right]=i\theta^{\mu\nu},
\end{equation}
with $\theta^{\mu\nu}$ being an antisymmetric tensor of second order not depending on the spacetime coordinates.
To formulate field theories on such a spacetime, the moyal star product can be used, which maps products
of fields depending on noncommuting coordinates to an extended expression containing the fields depending
on usual coordinates,

\begin{eqnarray}
\left.\varphi(\hat x)\psi(\hat x)=\varphi(x)\ast \psi(x)=\exp\left(\frac{i}{2}\theta^{\mu\nu}
\frac{\partial}{\partial x^\mu}\frac{\partial}{\partial y^\nu}\right)\varphi(x)\psi(y)\right|_{y \rightarrow x}
=\varphi(x)\psi(x)+\frac{i}{2}\theta^{\mu\nu}\partial_\mu \varphi(x) \partial_\nu \psi(x)
+\mathcal{O}\left(\theta^2\right).
\label{moyal_star_product}
\end{eqnarray}
Thus the moyal star product enables the possibility to represent field theories on noncommutative spacetime
as field theories with generalized expressions on usual spacetime. Performing the transition $x^\mu \rightarrow \hat x^\mu$
with respect to the conformally invariant gravity action ($\ref{usual_conformal_gravity_tetrad}$) and replacing
the thus obtained products of fields depending on noncommuting coordinates by the moyal star product according
to ($\ref{moyal_star_product}$) leads to a generalized action, which is not invariant under conformal transformations
anymore. The reason is that within the conformal transformations ($\ref{conformal_transformations_tetrad}$), the usual
coordinates also have to be replaced by the noncommuting ones leading to generalized transformation rules.
To maintain invariance under these generalized conformal transformations, generalized fields as well as a
generalized gauge parameter have to be introduced, which can be represented by the usual fields and the
usual gauge parameter by using Seiberg-Witten maps. The concept of Seiberg-Witten maps was originally introduced
in \cite{Seiberg:1999vs}. Accordingly, in the conformal gravity action ($\ref{usual_conformal_gravity_tetrad}$)
not only the usual products have to be replaced by moyal star products as representation of products between
fields depending on noncommutative coordinates, but also the fields itself have to be
replaced by generalized fields, $f\left(\hat x\right) \rightarrow \hat f\left(\hat x\right)$,
where the noncommutative fields $\hat f$ are related to the usual fields $f$ by Seiberg-Witten maps.
In case of conformal transformations the intricacy arises that in the usual case ($\ref{conformal_transformations_tetrad}$)
multiplying of the transformation operator from the left-hand side and from the right-hand side is equal, but
in the noncommutative case this is not the case anymore, since the sign of the exponential in the moyal star product
($\ref{moyal_star_product}$) is inverted, if the order of the factors is inverted. Thus the transformation in the noncommutative case has to be defined to be performed from the right-hand side or from the left-hand side.
Conventionally, noncommutative transformations are performed from the left-hand side in the noncommutative case,
since for a Lie group, the $U(1)$ excepted, the transformation has to be performed this way. The transformation of the
adjoint field is then performed from the right-hand side. Accordingly, in this paper the conformal transformation
of $\phi$ and $E^\mu_m$, which is performed by the operator $e^{-\lambda}$, is assumed to be performed from the left-hand
side in the noncommutative case, whereas the conformal transformation of $E_\mu^m$ as inverse of $E^\mu_m$, which is
performed by the operator $e^{\lambda}$, is performed from the right-hand side in the noncommutative case. This definition
is in a certain sense analogue to the case of noncommutative Yang-Mills transformations and it maintains that
Seiberg-Witten maps can be found and that the Seiberg-Witten map of the tetrad field $E_\mu^m$ is isomorphic to the Seiberg-Witten map of the inverse tetrad field $E^\mu_m$ what will be considered below. Thus the noncommutative gauge
transformations of the usual fields are defined as follows:

\begin{eqnarray}
&\phi(\hat x)\quad &\rightarrow \quad e^{-\lambda(\hat x)}\phi(\hat x)=e^{-\lambda(x)}\ast \phi(x)
=\phi(x)-\lambda(x)\ast \phi(x)+\mathcal{O}\left(\lambda^2\right),\nonumber\\
&E^\mu_m(\hat x)\quad &\rightarrow \quad e^{-\lambda(\hat x)} E^\mu_m(\hat x)=e^{-\lambda(x)}\ast E^\mu_m(x)
=E^\mu_m(x)-\lambda(x)\ast E_\mu^m(x)+\mathcal{O}\left(\lambda^2\right),\nonumber\\
&E_\mu^m(\hat x)\quad &\rightarrow \quad E_\mu^m(\hat x) e^{\lambda(\hat x)}=E_\mu^m(x)\ast e^{\lambda(x)}
=E_\mu^m(x)+E_\mu^m(x)\ast \lambda(x)+\mathcal{O}\left(\lambda^2\right).
\label{transformation_star-product}
\end{eqnarray}
The replacement of the usual fields by the generalized fields leads to a generalized action corresponding
to ($\ref{usual_conformal_gravity_tetrad}$), which is of the following form:

\begin{eqnarray}
\mathcal{S}_C&=&\int d^4 x\ \hat E \ast \left[\frac{1}{2} \hat E^{\mu m}\ast \hat E^\nu_m
\ast \partial_\mu \hat \phi \ast \partial_\nu \hat \phi
+\frac{1}{12} \hat E^\mu_a \ast \hat E^\nu_b \ast \hat R_{\mu\nu}^{ab}\left(\hat E\right)\ast \hat \phi \ast \hat \phi
+\hat C_{\mu\nu\rho\sigma}\left(\hat E\right)\ast \hat C^{\mu\nu\rho\sigma}\left(\hat E\right)\right]\nonumber\\
&=&\int d^4 x\ \hat E \left[\frac{1}{2}\hat E^{\mu m}\ast \hat E^\nu_m
\ast \partial_\mu \hat \phi \ast \partial_\nu \hat \phi
+\frac{1}{12}\hat E^\mu_a \ast \hat E^\nu_b \ast \hat R_{\mu\nu}^{ab}\left(\hat E\right)\ast \hat \phi \ast \hat \phi
+\hat C_{\mu\nu\rho\sigma}\left(\hat E\right) \ast \hat C^{\mu\nu\rho\sigma}\left(\hat E\right)\right],
\label{noncommutative_action}
\end{eqnarray}
where has been used in the second line that the following integral relation holds for the moyal star product:

\begin{equation}
\int d^4 x \left[f \ast g \ast h\right]=\int d^4 x \left[f\cdot \left(g\ast h\right)\right].
\end{equation}
In general the Seiberg-Witten maps are determined by the condition that a usual gauge transformation acting
on the usual fields within the expressions of the generalized fields, which are defined by the Seiberg-Witten
maps, is equal to a noncommutative gauge transformation, where the generalized gauge parameter acts on the
generalized fields by presupposing the star product. In the special case of this paper, the conditions for the
Seiberg-Witten maps are accordingly of the following form:

\begin{equation}
\delta \hat \phi=\hat \delta \hat \phi,\quad
\delta \hat E^\mu_m=\hat \delta \hat E^\mu_m,\quad
\delta \hat E_\mu^m=\hat \delta \hat E_\mu^m,
\label{condition_Seiberg-Witten}
\end{equation}
where the noncommutative gauge transformations of the noncommutative fields are in analogy to
($\ref{transformation_star-product}$) defined as following:

\begin{eqnarray}
\hat \delta \hat \phi(x)&=&e^{-\hat \lambda(x)}\ast \hat \phi(x)-\hat \phi(x)=-\hat \lambda(x)\ast \hat \phi(x)
+\mathcal{O}\left(\lambda^2\right),\nonumber\\
\hat \delta \hat E^\mu_m(x)&=&e^{-\hat \lambda(x)}\ast \hat E^\mu_m(x)-\hat E^\mu_m(x)=-\hat \lambda(x)\ast \hat E^\mu_m(x)
+\mathcal{O}\left(\lambda^2\right),\nonumber\\
\hat \delta \hat E_\mu^m(x)&=&\hat E_\mu^m(x) \ast e^{\hat \lambda(x)}-\hat E_\mu^m(x)=\hat E_\mu^m(x)\ast \hat \lambda(x) 
+\mathcal{O}\left(\lambda^2\right).
\label{transformations_generalized_fields}
\end{eqnarray}
The conditions ($\ref{condition_Seiberg-Witten}$) containing the usual conformal transformation rule 
($\ref{conformal_transformations_tetrad}$) and the noncommutative transformation
rule ($\ref{transformations_generalized_fields}$) is fulfilled by the following Seiberg-Witten maps
for the scalar field $\hat \phi$, the tetrad field $\hat E_\mu^m$, the inverse tetrad field $\hat E^\mu_m$
and the conformal gauge transformation parameter $\hat \lambda$:

\begin{eqnarray}
&&\hat \phi = \phi,\quad
\hat E^\mu_m = E^\mu_m+\frac{i}{2\phi}\theta^{\rho\sigma}\partial_\rho \phi \partial_\sigma E^\mu_m
+\mathcal{O}\left(\theta^2\right),\quad
\hat E_\mu^m = E_\mu^m+\frac{i}{2\phi}\theta^{\rho\sigma}\partial_\rho \phi \partial_\sigma E_\mu^m
+\mathcal{O}\left(\theta^2\right),\quad \nonumber\\&&
\hat \lambda = \lambda-\frac{i}{2\phi}\theta^{\rho\sigma}\partial_\rho \lambda \partial_\sigma \phi
+\mathcal{O}\left(\theta^2\right).
\label{Seiberg-Witten}
\end{eqnarray}
By inserting these expressions ($\ref{Seiberg-Witten}$) to the conditions ($\ref{condition_Seiberg-Witten}$) and
showing that both sides of the equations are equal, the validity of the Seiberg-Witten maps ($\ref{Seiberg-Witten}$)
can be proved. For the Seiberg-Witten map referring to the scalar field $\phi$, the corresponding calculation
is the following:

\begin{equation}
-\lambda \phi=-\left(\lambda-\frac{i}{2\phi}\theta^{\rho\sigma}\partial_\rho \lambda \partial_\sigma \phi\right)
\ast \phi+\mathcal{O}\left(\theta^2\right)
\quad \Leftrightarrow \quad
-\lambda \phi=-\lambda \phi-\frac{i}{2}\theta^{\rho\sigma}\partial_\rho \lambda \partial_\sigma \phi
+\frac{i}{2}\theta^{\rho\sigma}\partial_\rho \lambda \partial_\sigma \phi
+\mathcal{O}\left(\theta^2\right)
\quad \Leftrightarrow \quad
-\lambda \phi=-\lambda \phi,
\label{Seiberg-Witten_phi}
\end{equation}
and for the Seiberg-Witten map referring to the inverse tetrad field $E_\mu^m$ the calculation is the following:

\begin{eqnarray}
&&-\lambda E^\mu_m+\frac{i\lambda}{2\phi}\theta^{\rho\sigma}\partial_\rho \phi \partial_\sigma E^\mu_m
-\frac{i}{2\phi}\theta^{\rho\sigma}\partial_\rho \left(\lambda \phi\right) \partial_\sigma E^\mu_m
-\frac{i}{2\phi}\theta^{\rho\sigma}\partial_\rho \phi \partial_\sigma\left(\lambda E^\mu_m\right)
+\mathcal{O}\left(\theta^2\right)\nonumber\\
&&=-\left(\lambda-\frac{i}{2\phi}\theta^{\rho\sigma}\partial_\rho \lambda \partial_\sigma \phi\right)\ast
\left(E^\mu_m+\frac{i}{2\phi}\theta^{\rho\sigma}\partial_\rho \phi \partial_\sigma E^\mu_m\right)
+\mathcal{O}\left(\theta^2\right)
\nonumber\\ \quad \Leftrightarrow \quad
&&-\lambda E^\mu_m+\frac{i\lambda}{2\phi}\theta^{\rho\sigma}\partial_\rho \phi \partial_\sigma E^\mu_m
-\frac{i}{2}\theta^{\rho\sigma}\partial_\rho \lambda \partial_\sigma E^\mu_m
-\frac{i\lambda}{2\phi}\theta^{\rho\sigma}\partial_\rho \phi \partial_\sigma E^\mu_m
-\frac{iE^\mu_m}{2\phi}\theta^{\rho\sigma}\partial_\rho \phi \partial_\sigma \lambda
-\frac{i\lambda}{2\phi}\theta^{\rho\sigma}\partial_\rho \phi \partial_\sigma E^\mu_m
+\mathcal{O}\left(\theta^2\right)
\nonumber\\
&&=-\lambda E^\mu_m-\frac{i}{2}\theta^{\rho\sigma}\partial_\rho \lambda \partial_\sigma E^\mu_m
+\frac{iE^\mu_m}{2\phi}\theta^{\rho\sigma}\partial_\rho \lambda \partial_\sigma \phi
-\frac{i\lambda}{2\phi}\theta^{\rho\sigma}\partial_\rho \phi \partial_\sigma E^\mu_m
+\mathcal{O}\left(\theta^2\right)
\nonumber\\ \quad \Leftrightarrow \quad
&&-\lambda E^\mu_m-\frac{i\lambda}{2\phi}\theta^{\rho\sigma}\partial_\rho \phi \partial_\sigma E^\mu_m
-\frac{i}{2}\theta^{\rho\sigma}\partial_\rho \lambda \partial_\sigma E^\mu_m
-\frac{iE^\mu_m}{2\phi}\theta^{\rho\sigma}\partial_\rho \phi \partial_\sigma \lambda
+\mathcal{O}\left(\theta^2\right)
\nonumber\\
&&=-\lambda E^\mu_m-\frac{i}{2}\theta^{\rho\sigma}\partial_\rho \lambda \partial_\sigma E^\mu_m
+\frac{iE^\mu_m}{2\phi}\theta^{\rho\sigma}\partial_\rho \lambda \partial_\sigma \phi
-\frac{i\lambda}{2\phi}\theta^{\rho\sigma}\partial_\rho \phi \partial_\sigma E^\mu_m
+\mathcal{O}\left(\theta^2\right)
\nonumber\\ \quad \Leftrightarrow \quad
&&-\lambda E^\mu_m-\frac{i\lambda}{2\phi}\theta^{\rho\sigma}\partial_\rho \phi \partial_\sigma E^\mu_m
-\frac{i}{2}\theta^{\rho\sigma}\partial_\rho \lambda \partial_\sigma E^\mu_m
-\frac{iE^\mu_m}{2\phi}\theta^{\rho\sigma}\partial_\rho \phi \partial_\sigma \lambda
+\mathcal{O}\left(\theta^2\right)
\nonumber\\
&&=-\lambda E^\mu_m-\frac{i\lambda}{2\phi}\theta^{\rho\sigma}\partial_\rho \phi \partial_\sigma E^\mu_m
-\frac{i}{2}\theta^{\rho\sigma}\partial_\rho \lambda \partial_\sigma E^\mu_m
-\frac{iE^\mu_m}{2\phi}\theta^{\rho\sigma}\partial_\rho \phi \partial_\sigma \lambda
+\mathcal{O}\left(\theta^2\right).
\label{Seiberg-Witten_E}
\end{eqnarray}
The calculation yielding the proof of the validity of the Seiberg-Witten map of the tetrad field $E_\mu^m$ is not
listed here, since it is completely analogous to the one of $E^\mu_m$ with different signs at the beginning. 
Because of ($\ref{Seiberg-Witten_phi}$) and ($\ref{Seiberg-Witten_E}$) it is clear that the noncommutative gauge
transformations $\hat \delta$ fulfil also the consistency condition, that the commutator of two gauge
transformations is again a gauge transformation, the trivial transformation in this case:

\begin{equation}
\hat \delta_\lambda \hat \delta_\kappa-\hat \delta_\kappa \hat \delta_\lambda=\hat \delta_{[\lambda,\kappa]}=0.
\end{equation}

\section{Generalized Dynamics induced by the Noncommutativity}

It is now possible to calculate the action ($\ref{noncommutative_action}$) as generalization of ($\ref{usual_conformal_gravity}$) preserving conformal invariance on noncommutative spacetime and
containing the generalized fields being related to the usual fields by ($\ref{Seiberg-Witten}$).
To obtain the elaborated expression for the action ($\ref{noncommutative_action}$), the several
appearing factors have to be determined. The determinant of the generalized tetrad field
containing the star product and the Seiberg-Witten map is calculated as following:

\begin{eqnarray}
\hat E&=&\frac{1}{24}\epsilon^{\mu\nu\rho\sigma}\epsilon_{abcd}\hat E_\mu^a \ast
\hat E_\nu^b \ast \hat E_\rho^{c} \ast \hat E_\sigma^d\nonumber\\
&=&\frac{1}{24}\epsilon^{\mu\nu\rho\sigma}\epsilon_{abcd} E_\mu^a E_\nu^b E_\rho^{c} E_\sigma^d
+\frac{1}{4}\epsilon^{\mu\nu\rho\sigma}\epsilon_{abcd}\left(\frac{i}{2}\theta^{\lambda\kappa} \partial_\lambda E_\mu^a
\partial_\kappa E_\nu^b\right) E_\rho^{c} E_\sigma^d
+\frac{1}{24}\epsilon^{\mu\nu\rho\sigma}\epsilon_{abcd}\left(\frac{i}{2\phi}\theta^{\lambda\kappa}\partial_\lambda
\phi \partial_\kappa E_\mu^m\right) E_\nu^b E_\rho^{c} E_\sigma^d+\mathcal{O}\left(\theta^2\right),\nonumber\\
&=&\frac{1}{24}\epsilon^{\mu\nu\rho\sigma}\epsilon_{abcd} E_\mu^a E_\nu^b E_\rho^{c} E_\sigma^d
+\frac{i}{2}\theta^{\lambda\kappa}\left[\frac{1}{4}\epsilon^{\mu\nu\rho\sigma}\epsilon_{abcd} \partial_\lambda E_\mu^a
\partial_\kappa E_\nu^b E_\rho^{c} E_\sigma^d
+\frac{1}{24\phi}\epsilon^{\mu\nu\rho\sigma}\epsilon_{abcd}\partial_\lambda
\phi \partial_\kappa E_\mu^m E_\nu^b E_\rho^{c} E_\sigma^d\right]+\mathcal{O}\left(\theta^2\right),\nonumber\\
&\equiv&E+\mathcal{E}\left(E,\phi,\theta\right)+\mathcal{O}\left(\theta^2\right),
\label{noncommutative_determinant}
\end{eqnarray}
where $\epsilon^{\mu\nu\rho\sigma}$ denotes the totally antisymmetric tensor of fourth grade and $\mathcal{E}\left(E,\phi,\theta\right)$ contains all additional terms arising from the star product
and the Seiberg-Witten map. To calculate the generalized Riemann tensor on noncommutative spacetime
depending on the generalized tetrad field, $\hat R_{\mu\nu}^{ab}\left(\hat E\right)$, the
generalized connection $\hat \omega_\mu^{ab}\left(\hat E\right)$ has to be determined
first. Again this is done by expressing the connection in terms of the tetrad field,
replacing the usual products by moyal star products and replacing the tetrad
field by the generalized tetrad field and leads to

\begin{eqnarray}
\hat \omega_\mu^{ab}&=&2\hat E^{\nu a}\ast \partial_\mu \hat E_\nu^b
-2\hat E^{\nu b}\ast \partial_\mu \hat E_\nu^a
-2\hat E^{\nu a}\ast \partial_\nu \hat E_\mu^b+2\hat E^{\nu b}\ast \partial_\nu \hat E_\mu^a
+\hat E_{\mu c}\ast \hat E^{\nu a}\ast \hat E^{\sigma b}\ast \partial_\sigma \hat E_\nu^c
-\hat E_{\mu c}\ast \hat E^{\nu a}\ast \hat E^{\sigma b}\ast \partial_\nu \hat E_\sigma^c\nonumber\\
&=&2 E^{\nu a}\partial_\mu E_\nu^b-2 E^{\nu b} \partial_\mu E_\nu^a
-2 E^{\nu a} \partial_\nu E_\mu^b+2 E^{\nu b} \partial_\nu E_\mu^a
+E_{\mu c} E^{\nu a} E^{\sigma b} \partial_\sigma E_\nu^c
-E_{\mu c} E^{\nu a} E^{\sigma b} \partial_\nu E_\sigma^c\nonumber\\
&&+\frac{i}{2}\theta^{\lambda\kappa}\left[2 \partial_\lambda E^{\nu a} \partial_\kappa \partial_\mu E_\nu^b
-2 \partial_\lambda E^{\nu b} \partial_\kappa \partial_\mu E_\nu^a
-2 \partial_\lambda E^{\nu a} \partial_\kappa \partial_\nu E_\mu^b
+2 \partial_\lambda E^{\nu b} \partial_\kappa \partial_\nu E_\mu^a
\right.\nonumber\\&&\left.
+\partial_\lambda E_{\mu c} \partial_\kappa E^{\nu a} E^{\sigma b} \partial_\sigma E_\nu^c
+\partial_\lambda E_{\mu c} E^{\nu a} \partial_\kappa E^{\sigma b} \partial_\sigma E_\nu^c
+\partial_\lambda E_{\mu c} E^{\nu a} E^{\sigma b} \partial_\kappa \partial_\sigma E_\nu^c
\right.\nonumber\\&&\left.
+E_{\mu c} \partial_\lambda E^{\nu a} \partial_\kappa E^{\sigma b} \partial_\sigma E_\nu^c
+E_{\mu c} \partial_\lambda E^{\nu a} E^{\sigma b} \partial_\kappa \partial_\sigma E_\nu^c
+E_{\mu c} E^{\nu a} \partial_\lambda E^{\sigma b} \partial_\kappa \partial_\sigma E_\nu^c
\right.\nonumber\\&&\left.
-\partial_\lambda E_{\mu c} \partial_\kappa E^{\nu a} E^{\sigma b} \partial_\nu E_\sigma^c
-\partial_\lambda E_{\mu c} E^{\nu a} \partial_\kappa E^{\sigma b} \partial_\nu E_\sigma^c
-\partial_\lambda E_{\mu c} E^{\nu a} E^{\sigma b} \partial_\kappa \partial_\nu E_\sigma^c
\right.\nonumber\\&&\left.
-E_{\mu c} \partial_\lambda E^{\nu a} \partial_\kappa E^{\sigma b} \partial_\nu E_\sigma^c
-E_{\mu c} \partial_\lambda E^{\nu a} E^{\sigma b} \partial_\kappa \partial_\nu E_\sigma^c
-E_{\mu c} E^{\nu a} \partial_\lambda E^{\sigma b} \partial_\kappa \partial_\nu E_\sigma^c
\right.\nonumber\\&&\left.
+\frac{2}{\phi}\partial_\lambda \phi \partial_\kappa E^{\nu a} \partial_\mu E_\nu^b
-\frac{2\partial_\mu \phi}{\phi^2} \partial_\lambda \phi E^{\nu a} \partial_\kappa E_\nu^b
+\frac{2}{\phi}\partial_\mu \partial_\lambda \phi E^{\nu a} \partial_\kappa E_\nu^b
+\frac{2}{\phi}\partial_\lambda \phi E^{\nu a} \partial_\mu \partial_\kappa E_\nu^b
\right.\nonumber\\&&\left.
-\frac{2}{\phi}\partial_\lambda \phi \partial_\kappa E^{\nu b} \partial_\mu E_\nu^a
+\frac{2\partial_\mu \phi}{\phi^2} \partial_\lambda \phi E^{\nu b} \partial_\kappa E_\nu^a
-\frac{2}{\phi}\partial_\mu \partial_\lambda \phi E^{\nu b} \partial_\kappa E_\nu^a
-\frac{2}{\phi}\partial_\lambda \phi E^{\nu b} \partial_\mu \partial_\kappa E_\nu^a
\right.\nonumber\\&&\left.
-\frac{2}{\phi}\partial_\lambda \phi \partial_\kappa E^{\nu a} \partial_\nu E_\mu^b
+\frac{2\partial_\nu \phi}{\phi^2} \partial_\lambda \phi E^{\nu a} \partial_\kappa E_\mu^b
-\frac{2}{\phi}\partial_\nu \partial_\lambda \phi E^{\nu a} \partial_\kappa E_\mu^b
-\frac{2}{\phi}\partial_\lambda \phi E^{\nu a} \partial_\nu \partial_\kappa E_\mu^b
\right.\nonumber\\&&\left.
+\frac{2}{\phi}\partial_\lambda \phi \partial_\kappa E^{\nu b} \partial_\nu E_\mu^a
-\frac{2\partial_\nu \phi}{\phi^2} \partial_\lambda \phi E^{\nu b} \partial_\kappa E_\mu^a
+\frac{2}{\phi}\partial_\nu \partial_\lambda \phi E^{\nu b} \partial_\kappa E_\mu^a
+\frac{2}{\phi}\partial_\lambda \phi E^{\nu b} \partial_\nu \partial_\kappa E_\mu^a
\right.\nonumber\\&&\left.
+\frac{1}{\phi}\partial_\lambda \phi \partial_\kappa E_{\mu c}
E^{\nu a} E^{\sigma b} \partial_\sigma E_\nu^c
+\frac{1}{\phi}\partial_\lambda \phi E_{\mu c}
\partial_\kappa E^{\nu a} E^{\sigma b} \partial_\sigma E_\nu^c
+\frac{1}{\phi} \partial_\lambda \phi E_{\mu c} E^{\nu a}
\partial_\kappa E^{\sigma b} \partial_\sigma E_\nu^c
\right.\nonumber\\&&\left.
-\frac{\partial_\sigma \phi}{\phi^2} \partial_\lambda \phi E_{\mu c} E^{\nu a} E^{\sigma b}
\partial_\kappa E_\nu^c
+\frac{1}{\phi}\partial_\sigma \partial_\lambda \phi E_{\mu c} E^{\nu a} E^{\sigma b}
\partial_\kappa E_\nu^c
+\frac{1}{\phi}\partial_\lambda \phi E_{\mu c} E^{\nu a} E^{\sigma b}
\partial_\sigma \partial_\kappa E_\nu^c
\right.\nonumber\\&&\left.
-\frac{1}{\phi}\partial_\lambda \phi 
\partial_\kappa E_{\mu c} E^{\nu a} E^{\sigma b} \partial_\nu E_\sigma^c
-\frac{1}{\phi} \partial_\lambda \phi E_{\mu c}
\partial_\kappa E^{\nu a} E^{\sigma b} \partial_\nu E_\sigma^c
-\frac{1}{\phi} \partial_\lambda \phi E_{\mu c} E^{\nu a}
\partial_\kappa E^{\sigma b} \partial_\nu E_\sigma^c
\right.\nonumber\\&&\left.
+\frac{\partial_\nu \phi}{\phi^2}\partial_\lambda \phi E_{\mu c} E^{\nu a} E^{\sigma b}
\partial_\kappa E_\sigma^c
-\frac{1}{\phi}\partial_\nu \partial_\lambda \phi  E_{\mu c} E^{\nu a} E^{\sigma b}
\partial_\kappa E_\sigma^c
-\frac{1}{\phi}\partial_\lambda \phi E_{\mu c} E^{\nu a} E^{\sigma b}
\partial_\nu \partial_\kappa E_\sigma^c\right]+\mathcal{O}\left(\theta^2\right)\nonumber\\
&\equiv&\omega_\mu^{ab}\left(E\right)+\Omega_\mu^{ab}\left(E,\phi,\theta\right)+\mathcal{O}\left(\theta^2\right),
\label{noncommutative_connection}
\end{eqnarray}
where in the last line all additional terms are contained in the extension term $\Omega_\mu^{ab}\left(E,\phi,\theta\right)$,
which depends because of ($\ref{Seiberg-Witten}$) also on the scalar field $\phi$. By using ($\ref{noncommutative_connection}$),
the corresponding generalized Riemann tensor depending on the generalized tetrad field $\hat R_{\mu\nu}^{ab}\left(\hat E\right)$
can be calculated, 

\begin{eqnarray}
\hat R_{\mu\nu}^{ab}&=&\partial_\mu \hat \omega_\nu^{ab}\left(\hat E\right)
-\partial_\nu \hat \omega_\mu^{ab}\left(\hat E\right)
+\hat \omega_\mu^{ac}\left(\hat E\right) \ast \hat \omega_\nu^{cb}\left(\hat E\right)
-\hat \omega_\nu^{ac}\left(\hat E\right)\ast \hat \omega_\mu^{cb}\left(\hat E\right)\nonumber\\
&=&\partial_\mu \left[\omega_\nu^{ab}\left(E\right)+\Omega_\nu^{ab}\left(E,\phi,\theta\right)\right]
-\partial_\nu \left[\omega_\mu^{ab}\left(E\right)+\Omega_\mu^{ab}\left(E,\phi,\theta\right)\right]
\nonumber\\&&
+\left[\omega_\mu^{ac}\left(E\right)+\Omega_\mu^{ac}\left(E,\phi,\theta\right)\right] \ast
\left[\omega_\nu^{cb}\left(E\right)+\Omega_\nu^{cb}\left(E,\phi,\theta\right)\right]
\nonumber\\&&
-\left[\omega_\nu^{ac}\left(E\right)+\Omega_\nu^{ac}\left(E,\phi,\theta\right)\right]
\ast \left[\omega_\mu^{cb}\left(E\right)+\Omega_\mu^{cb}\left(E,\phi,\theta\right)\right]+\mathcal{O}\left(\theta^2\right)\nonumber\\
&=&\partial_\mu \omega_\nu^{ab}\left(E\right)-\partial_\nu \omega_\mu^{ab}\left(E\right)
+\omega_\mu^{ac}\left(E\right)\omega_\nu^{cb}\left(E\right)-\omega_\nu^{ac}\left(E\right)\omega_\mu^{cb}\left(E\right)
+\partial_\mu \Omega_\nu^{ab}\left(E,\phi,\theta\right)-\partial_\nu \Omega_\mu^{ab}\left(E,\phi,\theta\right)
\nonumber\\&&
+\frac{i}{2}\theta^{\lambda\kappa}\partial_\lambda \omega_\mu^{ac}\left(E\right) \partial_\kappa \omega_\nu^{cb}\left(E\right)
+\omega_\mu^{ac}\left(E\right) \Omega_\nu^{cb}\left(E,\phi,\theta\right)
+\Omega_\mu^{ac}\left(E,\phi,\theta\right) \omega_\nu^{cb}\left(E\right)
\nonumber\\&&
-\frac{i}{2}\theta^{\lambda\kappa}\partial_\lambda \omega_\nu^{ac}\left(E\right) \partial_\kappa \omega_\mu^{cb}\left(E\right)
-\omega_\nu^{ac}\left(E\right)\Omega_\mu^{cb}\left(E,\phi,\theta\right)
-\Omega_\nu^{ac}\left(E,\phi,\theta\right)\omega_\mu^{cb}\left(E\right)+\mathcal{O}\left(\theta^2\right)
\nonumber\\
&\equiv&R_{\mu\nu}^{ab}\left(E\right)+\mathcal{R}_{\mu\nu}^{ab}\left(E,\phi,\theta\right)+\mathcal{O}\left(\theta^2\right),
\label{noncommutative_Riemann_tensor}
\end{eqnarray}
where all additional terms are contained in the extension $\mathcal{R}_{\mu\nu}^{ab}\left(E,\phi,\theta\right)$, which again
also depends on the scalar field. With the generalized expression for the Riemann tensor ($\ref{noncommutative_Riemann_tensor}$)
the generalized expression for the conformal Weyl tensor depending on the generalized tetrad field
$\hat C_{\mu\nu\rho\sigma}\left(\hat E\right)$ can be calculated,

\begin{eqnarray}
\hat C_{\mu\nu\rho\sigma}&=&\hat R_{\mu\nu\rho\sigma}-
\left(\hat g_{\mu\rho}\ast \hat R_{\sigma\nu}
-\hat g_{\mu\sigma}\ast \hat R_{\rho\nu}-\hat g_{\nu\rho}\ast \hat R_{\sigma\mu}
+\hat g_{\nu\sigma}\ast \hat R_{\rho\mu}\right)+\frac{1}{3}
\left(\hat R\ast \hat g_{\mu\rho}\ast \hat g_{\sigma\nu}-\hat R \ast \hat g_{\mu\sigma}\ast \hat g_{\rho\nu}\right)\nonumber\\
&=&\hat E_{\nu a} \ast \hat E_{\sigma b} \ast \hat R_{\mu\rho}^{ab}-
\left(\hat E_\mu^a \ast \hat E_{\rho a}\ast \hat E_{\nu b}\ast \hat E^\tau_c \ast \hat R_{\sigma\tau}^{bc}
-\hat E_\mu^a \ast \hat E_{\sigma a} \ast \hat E_{\nu b}\ast \hat E^\tau_c \ast \hat R_{\rho\tau}^{bc} 
-\hat E_\nu^a \ast \hat E_{\rho a} \ast \hat E_{\mu b}\ast \hat E^\tau_c \ast \hat R_{\sigma\tau}^{bc}
\right.\nonumber\\&&\left.
+\hat E_\nu^a \ast \hat E_{\sigma a} \ast \hat E_{\mu b}\ast \hat E^\tau_c \ast \hat R_{\rho\tau}^{bc}\right)
+\frac{1}{3}\left(\hat E^\tau_b \ast \hat E^\upsilon_c \ast \hat R_{\tau\upsilon}^{bc} \ast \hat E_\mu^a \ast
\hat E_{\rho a}\ast \hat E_\sigma^a \ast \hat E_{\nu a}
-\hat E^\tau_b \ast \hat E^\upsilon_c \ast \hat R_{\tau\upsilon}^{bc} \ast \hat E_\mu^a \ast
\hat E_{\sigma a}\ast \hat E_\rho^a \ast E_{\nu a}\right)\nonumber\\
&=&E_{\nu a} E_{\sigma b} R_{\mu\rho}^{ab}-
\left(E_\mu^a E_{\rho a} E_{\nu b} E^\tau_c R_{\sigma\tau}^{bc}
-E_\mu^a E_{\sigma a} E_{\nu b} E^\tau_c R_{\rho\tau}^{bc} 
-E_\nu^a E_{\rho a} E_{\mu b} E^\tau_c R_{\sigma\tau}^{bc}
+E_\nu^a E_{\sigma a} E_{\mu b} E^\tau_c R_{\rho\tau}^{bc}\right)
\nonumber\\&&
+\frac{1}{3}\left(E^\tau_b E^\upsilon_c R_{\tau\upsilon}^{bc} E_\mu^a E_{\rho a} E_\sigma^a E_{\nu a}
-E^\tau_b E^\upsilon_c R_{\tau\upsilon}^{bc} E_\mu^a E_{\sigma a} E_\rho^a E_{\nu a}\right)
+\mathcal{A}_{\mu\nu\rho\sigma}\left(E,\theta\right)+\mathcal{B}_{\mu\nu\rho\sigma}\left(E,\phi,\theta\right)+\mathcal{O}\left(\theta^2\right)
\nonumber\\&\equiv&C_{\mu\nu\rho\sigma}\left(E\right)+\mathcal{C}_{\mu\nu\rho\sigma}\left(E,\phi,\theta\right)
+\mathcal{O}\left(\theta^2\right),
\label{generalized_conformal_tensor}
\end{eqnarray}
where $\mathcal{C}_{\mu\nu\rho\sigma}\left(E,\phi,\theta\right)=\mathcal{A}_{\mu\nu\rho\sigma}\left(E,\theta\right)
+\mathcal{B}_{\mu\nu\rho\sigma}\left(E,\phi,\theta\right)$ and thus contains all additional terms,
$\mathcal{A}_{\mu\nu\rho\sigma}\left(E,\theta\right)$ contains all additional terms arising from the the star product  
and $\mathcal{B}_{\mu\nu\rho\sigma}\left(E,\phi,\theta\right)$ contains all additional terms arising from the
Seiberg-Witten map. Accordingly the tensors $\mathcal{A}_{\mu\nu\rho\sigma}\left(E,\theta\right)$
and $\mathcal{B}_{\mu\nu\rho\sigma}\left(E,\phi,\theta\right)$ are defined as

\begin{eqnarray}
\mathcal{A}_{\mu\nu\rho\sigma}
&=&\frac{i}{2}\theta^{\lambda\kappa}\left[ \partial_\lambda E_{\nu a} \partial_\kappa E_{\sigma b} R_{\mu\rho}^{ab}
+\partial_\lambda E_{\nu a} E_{\sigma b} \partial_\kappa R_{\mu\rho}^{ab}
+E_{\nu a} \partial_\lambda E_{\sigma b} \partial_\kappa R_{\mu\rho}^{ab}
\right.\nonumber\\&&\left.
-\left(
\partial_\lambda E_\mu^a \partial_\kappa E_{\rho a} E_{\nu b} E^\tau_c R_{\sigma\tau}^{bc}
+\partial_\lambda E_\mu^a E_{\rho a} \partial_\kappa E_{\nu b} E^\tau_c R_{\sigma\tau}^{bc}
+\partial_\lambda E_\mu^a E_{\rho a} E_{\nu b} \partial_\kappa E^\tau_c R_{\sigma\tau}^{bc}
+\partial_\lambda E_\mu^a E_{\rho a} E_{\nu b} E^\tau_c \partial_\kappa R_{\sigma\tau}^{bc}
\right.\right.\nonumber\\&&\left.\left.
+E_\mu^a \partial_\lambda E_{\rho a} \partial_\kappa E_{\nu b} E^\tau_c R_{\sigma\tau}^{bc}
+E_\mu^a \partial_\lambda E_{\rho a} E_{\nu b} \partial_\kappa E^\tau_c R_{\sigma\tau}^{bc}
+E_\mu^a \partial_\lambda E_{\rho a} E_{\nu b} E^\tau_c \partial_\kappa R_{\sigma\tau}^{bc}
\right.\right.\nonumber\\&&\left.\left.
+E_\mu^a E_{\rho a} \partial_\lambda E_{\nu b} \partial_\kappa E^\tau_c R_{\sigma\tau}^{bc}
+E_\mu^a E_{\rho a} \partial_\lambda E_{\nu b} E^\tau_c \partial_\kappa R_{\sigma\tau}^{bc}
+E_\mu^a E_{\rho a} E_{\nu b} \partial_\lambda E^\tau_c \partial_\kappa R_{\sigma\tau}^{bc}
\right.\right.\nonumber\\&&\left.\left.
-\partial_\lambda E_\mu^a \partial_\kappa E_{\sigma a} E_{\nu b} E^\tau_c R_{\rho\tau}^{bc} 
-\partial_\lambda E_\mu^a E_{\sigma a} \partial_\kappa E_{\nu b} E^\tau_c R_{\rho\tau}^{bc} 
-\partial_\lambda E_\mu^a E_{\sigma a} E_{\nu b} \partial_\kappa E^\tau_c R_{\rho\tau}^{bc} 
-\partial_\lambda E_\mu^a E_{\sigma a} E_{\nu b} E^\tau_c \partial_\kappa R_{\rho\tau}^{bc} 
\right.\right.\nonumber\\&&\left.\left.
-E_\mu^a \partial_\lambda E_{\sigma a} \partial_\kappa E_{\nu b} E^\tau_c R_{\rho\tau}^{bc}
-E_\mu^a \partial_\lambda E_{\sigma a} E_{\nu b} \partial_\kappa E^\tau_c R_{\rho\tau}^{bc} 
-E_\mu^a \partial_\lambda E_{\sigma a} E_{\nu b} E^\tau_c \partial_\kappa R_{\rho\tau}^{bc} 
\right.\right.\nonumber\\&&\left.\left.
-E_\mu^a E_{\sigma a} \partial_\lambda E_{\nu b} \partial_\kappa E^\tau_c R_{\rho\tau}^{bc} 
-E_\mu^a E_{\sigma a} \partial_\lambda E_{\nu b} E^\tau_c \partial_\kappa R_{\rho\tau}^{bc} 
-E_\mu^a E_{\sigma a} E_{\nu b} \partial_\lambda E^\tau_c \partial_\kappa R_{\rho\tau}^{bc}
\right.\right.\nonumber\\&&\left.\left.
-\partial_\lambda E_\nu^a \partial_\kappa E_{\rho a} E_{\mu b} E^\tau_c R_{\sigma\tau}^{bc}
-\partial_\lambda E_\nu^a E_{\rho a} \partial_\kappa E_{\mu b} E^\tau_c R_{\sigma\tau}^{bc}
-\partial_\lambda E_\nu^a E_{\rho a} E_{\mu b} \partial_\kappa E^\tau_c R_{\sigma\tau}^{bc}
-\partial_\lambda E_\nu^a E_{\rho a} E_{\mu b} E^\tau_c \partial_\kappa R_{\sigma\tau}^{bc}
\right.\right.\nonumber\\&&\left.\left.
-E_\nu^a \partial_\lambda E_{\rho a} \partial_\kappa E_{\mu b} E^\tau_c R_{\sigma\tau}^{bc}
-E_\nu^a \partial_\lambda E_{\rho a} E_{\mu b} \partial_\kappa E^\tau_c R_{\sigma\tau}^{bc}
-E_\nu^a \partial_\lambda E_{\rho a} E_{\mu b} E^\tau_c \partial_\kappa R_{\sigma\tau}^{bc}
\right.\right.\nonumber\\&&\left.\left.
-E_\nu^a E_{\rho a} \partial_\lambda E_{\mu b} \partial_\kappa E^\tau_c R_{\sigma\tau}^{bc}
-E_\nu^a E_{\rho a} \partial_\lambda E_{\mu b} E^\tau_c \partial_\kappa R_{\sigma\tau}^{bc}
-E_\nu^a E_{\rho a} E_{\mu b} \partial_\lambda E^\tau_c \partial_\kappa R_{\sigma\tau}^{bc}
\right.\right.\nonumber\\&&\left.\left.
+\partial_\lambda E_\nu^a \partial_\kappa E_{\sigma a} E_{\mu b} E^\tau_c R_{\rho\tau}^{bc}
+\partial_\lambda E_\nu^a E_{\sigma a} \partial_\kappa E_{\mu b} E^\tau_c R_{\rho\tau}^{bc}
+\partial_\lambda E_\nu^a E_{\sigma a} E_{\mu b} \partial_\kappa E^\tau_c R_{\rho\tau}^{bc}
+\partial_\lambda E_\nu^a E_{\sigma a} E_{\mu b} E^\tau_c \partial_\kappa R_{\rho\tau}^{bc}
\right.\right.\nonumber\\&&\left.\left.
+E_\nu^a \partial_\lambda E_{\sigma a} \partial_\kappa E_{\mu b} E^\tau_c R_{\rho\tau}^{bc}
+E_\nu^a \partial_\lambda E_{\sigma a} E_{\mu b} \partial_\kappa E^\tau_c R_{\rho\tau}^{bc}
+E_\nu^a \partial_\lambda E_{\sigma a} E_{\mu b} E^\tau_c \partial_\kappa R_{\rho\tau}^{bc}
\right.\right.\nonumber\\&&\left.\left.
+E_\nu^a E_{\sigma a} \partial_\lambda E_{\mu b} \partial_\kappa E^\tau_c R_{\rho\tau}^{bc}
+E_\nu^a E_{\sigma a} \partial_\lambda E_{\mu b} E^\tau_c \partial_\kappa R_{\rho\tau}^{bc}
+E_\nu^a E_{\sigma a} E_{\mu b} \partial_\lambda E^\tau_c \partial_\kappa R_{\rho\tau}^{bc}\right)
\right.\nonumber\\&&\left.
+\frac{1}{3}
\left(\partial_\lambda E^\tau_b \partial_\kappa E^\upsilon_c R_{\tau\upsilon}^{bc} E_\mu^a E_{\rho a} E_\sigma^a E_{\nu a}
+\partial_\lambda E^\tau_b E^\upsilon_c \partial_\kappa R_{\tau\upsilon}^{bc} E_\mu^a E_{\rho a} E_\sigma^a E_{\nu a}
+\partial_\lambda E^\tau_b E^\upsilon_c R_{\tau\upsilon}^{bc} \partial_\kappa E_\mu^a E_{\rho a} E_\sigma^a E_{\nu a}
\right.\right.\nonumber\\&&\left.\left.
+\partial_\lambda E^\tau_b E^\upsilon_c R_{\tau\upsilon}^{bc} E_\mu^a \partial_\kappa E_{\rho a} E_\sigma^a E_{\nu a}
+\partial_\lambda E^\tau_b E^\upsilon_c R_{\tau\upsilon}^{bc} E_\mu^a E_{\rho a} \partial_\kappa E_\sigma^a E_{\nu a}
+\partial_\lambda E^\tau_b E^\upsilon_c R_{\tau\upsilon}^{bc} E_\mu^a E_{\rho a} E_\sigma^a \partial_\kappa E_{\nu a}
\right.\right.\nonumber\\&&\left.\left.
+E^\tau_b \partial_\lambda E^\upsilon_c \partial_\kappa R_{\tau\upsilon}^{bc} E_\mu^a E_{\rho a} E_\sigma^a E_{\nu a}
+E^\tau_b \partial_\lambda E^\upsilon_c R_{\tau\upsilon}^{bc} \partial_\kappa E_\mu^a E_{\rho a} E_\sigma^a E_{\nu a}
+E^\tau_b \partial_\lambda E^\upsilon_c R_{\tau\upsilon}^{bc} E_\mu^a \partial_\kappa E_{\rho a} E_\sigma^a E_{\nu a}
\right.\right.\nonumber\\&&\left.\left.
+E^\tau_b \partial_\lambda E^\upsilon_c R_{\tau\upsilon}^{bc} E_\mu^a E_{\rho a} \partial_\kappa E_\sigma^a E_{\nu a}
+E^\tau_b \partial_\lambda E^\upsilon_c R_{\tau\upsilon}^{bc} E_\mu^a E_{\rho a} E_\sigma^a \partial_\kappa E_{\nu a}
+E^\tau_b E^\upsilon_c \partial_\lambda R_{\tau\upsilon}^{bc} \partial_\kappa E_\mu^a E_{\rho a} E_\sigma^a E_{\nu a}
\right.\right.\nonumber\\&&\left.\left.
+E^\tau_b E^\upsilon_c \partial_\lambda R_{\tau\upsilon}^{bc} E_\mu^a \partial_\kappa E_{\rho a} E_\sigma^a E_{\nu a}
+E^\tau_b E^\upsilon_c \partial_\lambda R_{\tau\upsilon}^{bc} E_\mu^a E_{\rho a} \partial_\kappa E_\sigma^a E_{\nu a}
+E^\tau_b E^\upsilon_c \partial_\lambda R_{\tau\upsilon}^{bc} E_\mu^a E_{\rho a} E_\sigma^a \partial_\kappa E_{\nu a}
\right.\right.\nonumber\\&&\left.\left.
+E^\tau_b E^\upsilon_c R_{\tau\upsilon}^{bc} \partial_\lambda E_\mu^a \partial_\kappa E_{\rho a} E_\sigma^a E_{\nu a}
+E^\tau_b E^\upsilon_c R_{\tau\upsilon}^{bc} \partial_\lambda E_\mu^a E_{\rho a} \partial_\kappa E_\sigma^a E_{\nu a}
+E^\tau_b E^\upsilon_c R_{\tau\upsilon}^{bc} \partial_\lambda E_\mu^a E_{\rho a} E_\sigma^a \partial_\kappa E_{\nu a}
\right.\right.\nonumber\\&&\left.\left.
+E^\tau_b E^\upsilon_c R_{\tau\upsilon}^{bc} E_\mu^a \partial_\lambda E_{\rho a} \partial_\kappa E_\sigma^a E_{\nu a}
+E^\tau_b E^\upsilon_c R_{\tau\upsilon}^{bc} E_\mu^a \partial_\lambda E_{\rho a} E_\sigma^a \partial_\kappa E_{\nu a}
+E^\tau_b E^\upsilon_c R_{\tau\upsilon}^{bc} E_\mu^a E_{\rho a} \partial_\lambda E_\sigma^a \partial_\kappa E_{\nu a}
\right.\right.\nonumber\\&&\left.\left.
-\partial_\lambda E^\tau_b \partial_\kappa E^\upsilon_c R_{\tau\upsilon}^{bc} E_\mu^a E_{\sigma a} E_\rho^a E_{\nu a}
-\partial_\lambda E^\tau_b E^\upsilon_c \partial_\kappa R_{\tau\upsilon}^{bc} E_\mu^a E_{\sigma a} E_\rho^a E_{\nu a}
-\partial_\lambda E^\tau_b E^\upsilon_c R_{\tau\upsilon}^{bc} \partial_\kappa E_\mu^a E_{\sigma a} E_\rho^a E_{\nu a}
\right.\right.\nonumber\\&&\left.\left.
-\partial_\lambda E^\tau_b E^\upsilon_c R_{\tau\upsilon}^{bc} E_\mu^a \partial_\kappa E_{\sigma a} E_\rho^a E_{\nu a}
-\partial_\lambda E^\tau_b E^\upsilon_c R_{\tau\upsilon}^{bc} E_\mu^a E_{\sigma a} \partial_\kappa E_\rho^a E_{\nu a}
-\partial_\lambda E^\tau_b E^\upsilon_c R_{\tau\upsilon}^{bc} E_\mu^a E_{\sigma a} E_\rho^a \partial_\kappa E_{\nu a}
\right.\right.\nonumber\\&&\left.\left.
-E^\tau_b \partial_\lambda E^\upsilon_c \partial_\kappa R_{\tau\upsilon}^{bc} E_\mu^a E_{\sigma a} E_\rho^a E_{\nu a}
-E^\tau_b \partial_\lambda E^\upsilon_c R_{\tau\upsilon}^{bc} \partial_\kappa E_\mu^a E_{\sigma a} E_\rho^a E_{\nu a}
-E^\tau_b \partial_\lambda E^\upsilon_c R_{\tau\upsilon}^{bc} E_\mu^a \partial_\kappa E_{\sigma a} E_\rho^a E_{\nu a}
\right.\right.\nonumber\\&&\left.\left.
-E^\tau_b \partial_\lambda E^\upsilon_c R_{\tau\upsilon}^{bc} E_\mu^a E_{\sigma a} \partial_\kappa E_\rho^a E_{\nu a}
-E^\tau_b \partial_\lambda E^\upsilon_c R_{\tau\upsilon}^{bc} E_\mu^a E_{\sigma a} E_\rho^a \partial_\kappa E_{\nu a}
-E^\tau_b E^\upsilon_c \partial_\lambda R_{\tau\upsilon}^{bc} \partial_\kappa E_\mu^a E_{\sigma a} E_\rho^a E_{\nu a}
\right.\right.\nonumber\\&&\left.\left.
-E^\tau_b E^\upsilon_c \partial_\lambda R_{\tau\upsilon}^{bc} E_\mu^a \partial_\kappa E_{\sigma a} E_\rho^a E_{\nu a}
-E^\tau_b E^\upsilon_c \partial_\lambda R_{\tau\upsilon}^{bc} E_\mu^a E_{\sigma a} \partial_\kappa E_\rho^a E_{\nu a}
-E^\tau_b E^\upsilon_c \partial_\lambda R_{\tau\upsilon}^{bc} E_\mu^a E_{\sigma a} E_\rho^a \partial_\kappa E_{\nu a}
\right.\right.\nonumber\\&&\left.\left.
-E^\tau_b E^\upsilon_c R_{\tau\upsilon}^{bc} \partial_\lambda E_\mu^a \partial_\kappa E_{\sigma a} E_\rho^a E_{\nu a}
-E^\tau_b E^\upsilon_c R_{\tau\upsilon}^{bc} \partial_\lambda E_\mu^a E_{\sigma a} \partial_\kappa E_\rho^a E_{\nu a}
-E^\tau_b E^\upsilon_c R_{\tau\upsilon}^{bc} \partial_\lambda E_\mu^a E_{\sigma a} E_\rho^a \partial_\kappa E_{\nu a}
\right.\right.\nonumber\\&&\left.\left.
-E^\tau_b E^\upsilon_c R_{\tau\upsilon}^{bc} E_\mu^a \partial_\lambda E_{\sigma a} \partial_\kappa E_\rho^a E_{\nu a}
-E^\tau_b E^\upsilon_c R_{\tau\upsilon}^{bc} E_\mu^a \partial_\lambda E_{\sigma a} E_\rho^a \partial_\kappa E_{\nu a}
-E^\tau_b E^\upsilon_c R_{\tau\upsilon}^{bc} E_\mu^a E_{\sigma a} \partial_\lambda E_\rho^a \partial_\kappa E_{\nu a}
\right)\right],
\end{eqnarray}

\begin{eqnarray}
\mathcal{B}_{\mu\nu\rho\sigma}&=&\frac{i\theta^{\lambda\kappa}}{2\phi}\left(\partial_\lambda \phi \partial_\kappa E_{\nu a} E_{\sigma b} R_{\mu\rho}^{ab}+\partial_\lambda \phi E_{\nu a} \partial_\kappa
E_{\sigma b} R_{\mu\rho}^{ab}\right)+E_{\nu a} E_{\sigma b} \mathcal{R}_{\mu\rho}^{ab}\left(E,\phi,\theta\right)
\nonumber\\&&\
-\left[\frac{i\theta^{\lambda\kappa}}{2\phi}\left(\partial_\lambda \phi \partial_\kappa E_\mu^a E_{\rho a} E_{\nu b} E^\tau_c R_{\sigma\tau}^{bc}
+\partial_\lambda \phi E_\mu^a \partial_\kappa E_{\rho a} E_{\nu b} E^\tau_c R_{\sigma\tau}^{bc}
+\partial_\lambda \phi E_\mu^a E_{\rho a}\partial_\kappa E_{\nu b} E^\tau_c R_{\sigma\tau}^{bc}
+\partial_\lambda \phi E_\mu^a E_{\rho a} E_{\nu b} \partial_\kappa E^\tau_c R_{\sigma\tau}^{bc}
\right.\right.\nonumber\\&&\left.\left.
-\partial_\lambda \phi \partial_\kappa E_\mu^a E_{\sigma a} E_{\nu b} E^\tau_c R_{\rho\tau}^{bc} 
-\partial_\lambda \phi E_\mu^a \partial_\kappa E_{\sigma a} E_{\nu b} E^\tau_c R_{\rho\tau}^{bc} 
-\partial_\lambda \phi E_\mu^a E_{\sigma a} \partial_\kappa E_{\nu b} E^\tau_c R_{\rho\tau}^{bc} 
-\partial_\lambda \phi E_\mu^a E_{\sigma a} E_{\nu b} \partial_\kappa E^\tau_c R_{\rho\tau}^{bc}
\right.\right.\nonumber\\&&\left.\left.
-\partial_\lambda \phi \partial_\kappa E_\nu^a E_{\rho a} E_{\mu b} E^\tau_c R_{\sigma\tau}^{bc}
-\partial_\lambda \phi E_\nu^a \partial_\kappa E_{\rho a} E_{\mu b} E^\tau_c R_{\sigma\tau}^{bc}
-\partial_\lambda \phi E_\nu^a E_{\rho a}\partial_\kappa E_{\mu b} E^\tau_c R_{\sigma\tau}^{bc}
-\partial_\lambda \phi E_\nu^a E_{\rho a} E_{\mu b} \partial_\kappa E^\tau_c R_{\sigma\tau}^{bc}
\right.\right.\nonumber\\&&\left.\left.
+\partial_\lambda \phi \partial_\kappa E_\nu^a E_{\sigma a} E_{\mu b} E^\tau_c R_{\rho\tau}^{bc}
+\partial_\lambda \phi E_\nu^a \partial_\kappa E_{\sigma a} E_{\mu b} E^\tau_c R_{\rho\tau}^{bc}
+\partial_\lambda \phi E_\nu^a E_{\sigma a}\partial_\kappa E_{\mu b} E^\tau_c R_{\rho\tau}^{bc}
+\partial_\lambda \phi E_\nu^a E_{\sigma a} E_{\mu b}\partial_\kappa E^\tau_c R_{\rho\tau}^{bc}\right)
\right.\nonumber\\&&\left.
+E_\mu^a E_{\rho a} E_{\nu b} E^\tau_c \mathcal{R}_{\sigma\tau}^{bc}\left(E,\phi,\theta\right)
-E_\mu^a E_{\sigma a} E_{\nu b} E^\tau_c \mathcal{R}_{\rho\tau}^{bc}\left(E,\phi,\theta\right)
-E_\nu^a E_{\rho a} E_{\mu b} E^\tau_c \mathcal{R}_{\sigma\tau}^{bc}\left(E,\phi,\theta\right)
+E_\nu^a E_{\sigma a} E_{\mu b} E^\tau_c \mathcal{R}_{\rho\tau}^{bc}\left(E,\phi,\theta\right)\right]
\nonumber\\&&
+\frac{1}{3}\left[\frac{i\theta^{\lambda\kappa}}{2\phi}\left(\partial_\lambda \phi \partial_\kappa E^\tau_b E^\upsilon_c R_{\tau\upsilon}^{bc} E_\mu^a E_{\rho a} E_\sigma^a E_{\nu a}
+\partial_\lambda \phi E^\tau_b \partial_\kappa E^\upsilon_c R_{\tau\upsilon}^{bc} E_\mu^a E_{\rho a} E_\sigma^a E_{\nu a}
+\partial_\lambda \phi E^\tau_b E^\upsilon_c R_{\tau\upsilon}^{bc}\partial_\kappa E_\mu^a E_{\rho a} E_\sigma^a E_{\nu a}
\right.\right.\nonumber\\&&\left.\left.
+\partial_\lambda \phi E^\tau_b E^\upsilon_c R_{\tau\upsilon}^{bc} E_\mu^a \partial_\kappa E_{\rho a} E_\sigma^a E_{\nu a}
+\partial_\lambda \phi E^\tau_b E^\upsilon_c R_{\tau\upsilon}^{bc} E_\mu^a E_{\rho a} \partial_\kappa E_\sigma^a E_{\nu a}
+\partial_\lambda \phi E^\tau_b E^\upsilon_c R_{\tau\upsilon}^{bc} E_\mu^a E_{\rho a} E_\sigma^a \partial_\kappa E_{\nu a}
\right.\right.\nonumber\\&&\left.\left.
-\partial_\lambda \phi \partial_\kappa E^\tau_b E^\upsilon_c R_{\tau\upsilon}^{bc} E_\mu^a E_{\sigma a} E_\rho^a E_{\nu a}
-\partial_\lambda \phi E^\tau_b \partial_\kappa E^\upsilon_c R_{\tau\upsilon}^{bc} E_\mu^a E_{\sigma a} E_\rho^a E_{\nu a}
-\partial_\lambda \phi E^\tau_b E^\upsilon_c R_{\tau\upsilon}^{bc} \partial_\kappa E_\mu^a E_{\sigma a} E_\rho^a E_{\nu a}
\right.\right.\nonumber\\&&\left.\left.
-\partial_\lambda \phi E^\tau_b E^\upsilon_c R_{\tau\upsilon}^{bc} E_\mu^a \partial_\kappa E_{\sigma a} E_\rho^a E_{\nu a}
-\partial_\lambda \phi E^\tau_b E^\upsilon_c R_{\tau\upsilon}^{bc} E_\mu^a E_{\sigma a} \partial_\kappa E_\rho^a E_{\nu a}
-\partial_\lambda \phi E^\tau_b E^\upsilon_c R_{\tau\upsilon}^{bc} E_\mu^a E_{\sigma a} E_\rho^a
\partial_\kappa E_{\nu a}\right)
\right.\nonumber\\&&\left.
+E^\tau_b E^\upsilon_c \mathcal{R}_{\tau\upsilon}^{bc}\left(E,\phi,\theta\right) E_\mu^a E_{\rho a} E_\sigma^a E_{\nu a}
-E^\tau_b E^\upsilon_c \mathcal{R}_{\tau\upsilon}^{bc}\left(E,\phi,\theta\right) E_\mu^a E_{\sigma a} E_\rho^a E_{\nu a}\right].
\end{eqnarray}
Inserting the generalized determinant ($\ref{noncommutative_determinant}$), the generalized Riemann tensor ($\ref{noncommutative_Riemann_tensor}$) depending on the generalized connection defined in ($\ref{noncommutative_connection}$) 
and the generalized conformal tensor ($\ref{generalized_conformal_tensor}$) depending on
($\ref{noncommutative_Riemann_tensor}$) to the generalized conformally invariant gravity action ($\ref{noncommutative_action}$) yields the following expression for the generalized conformal gravity action:

\begin{eqnarray}
\mathcal{S}_C&=&\int d^4 x\left\{E \left[\frac{1}{2} E^{\mu m} E^\nu_m \partial_\mu \phi \partial_\nu \phi
+\frac{1}{12} E^\mu_a E^\nu_b R_{\mu\nu}^{ab}\left(E\right)\phi^2
+C_{\mu\nu\rho\sigma}\left(E\right)C^{\mu\nu\rho\sigma}\left(E\right)
+i\theta^{\lambda\kappa}\left(
\frac{1}{4} \partial_\lambda E^{\mu m} \partial_\kappa E^\nu_m \partial_\mu \phi \partial_\nu \phi
\right.\right.\right.\nonumber\\&&\left.\left.\left.+
\frac{1}{2}\partial_\lambda E^{\mu m} E^\nu_m \partial_\kappa \partial_\mu \phi \partial_\nu \phi
+\frac{1}{2}\partial_\lambda E^{\mu m} E^\nu_m \partial_\mu \phi \partial_\kappa \partial_\nu \phi
+\frac{1}{4} E^{\mu m} E^\nu_m \partial_\lambda \partial_\mu \phi \partial_\kappa \partial_\nu \phi
+\frac{1}{24} \partial_\lambda E^\mu_a \partial_\kappa E^\nu_b R_{\mu\nu}^{ab}\left(E\right) \phi^2
\right.\right.\right.\nonumber\\&&\left.\left.\left.
+\frac{1}{12} \partial_\lambda E^\mu_a E^\nu_b \partial_\kappa R_{\mu\nu}^{ab}\left(E\right) \phi^2
+\frac{1}{6} \partial_\lambda E^\mu_a E^\nu_b R_{\mu\nu}^{ab}\left(E\right) \phi \partial_\kappa \phi 
+\frac{1}{12} E^\mu_a E^\nu_b \partial_\lambda R_{\mu\nu}^{ab}\left(E\right) \phi \partial_\kappa \phi
+\frac{1}{24} E^\mu_a E^\nu_b R_{\mu\nu}^{ab}\left(E\right) \partial_\lambda \phi \partial_\kappa \phi
\right.\right.\right.\nonumber\\&&\left.\left.\left.
+\frac{1}{2\phi}\partial_\lambda \phi
\partial_\kappa E^{\mu m} E^\nu_m \partial_\mu \phi \partial_\nu \phi
+\frac{1}{12}\partial_\lambda \phi
\partial_\kappa E^\mu_a E^\nu_b R_{\mu\nu}^{ab}\left(E\right)\phi\right)
+\frac{1}{12} E^\mu_a E^\nu_b \mathcal{R}_{\mu\nu}^{ab}\left(E,\phi,\theta\right)\phi^2
+2C_{\mu\nu\rho\sigma}\left(E\right)\mathcal{C}^{\mu\nu\rho\sigma}\left(E,\phi,\theta\right)\right]
\right.\nonumber\\&&\left.
+\mathcal{E}\left(E,\phi,\theta\right)\left[\frac{1}{2} E^{\mu m} E^\nu_m \partial_\mu \phi \partial_\nu \phi
+\frac{1}{12} E^\mu_a E^\nu_b R_{\mu\nu}^{ab}\left(E\right)\phi^2
+C_{\mu\nu\rho\sigma}\left(E\right)C^{\mu\nu\rho\sigma}\left(E\right)\right]\right\}+\mathcal{O}\left(\theta^2\right).
\label{explicit_noncommutative_action}
\end{eqnarray}
Variation of this action ($\ref{explicit_noncommutative_action}$) with respect to $\phi$ and $E^\mu_m$, $\frac{\delta \mathcal{S}}{\delta \phi}$ and $\frac{\delta \mathcal{S}}{\delta E^\mu_a}$, yields the corresponding field equations:

\begin{eqnarray}
&&-\partial_\mu \left(E E^{\mu m} E^\nu_m \partial_\nu \phi\right)
+\frac{1}{6}E E^\mu_a E^\nu_b R_{\mu\nu}^{ab}\left(E\right)\phi
+i\theta^{\lambda\kappa}\left[
-\frac{1}{4}\partial_\mu \left(E \partial_\lambda E^{\mu m} \partial_\kappa E^\nu_m \partial_\nu \phi\right)
-\frac{1}{4}\partial_\nu \left(E \partial_\lambda E^{\mu m} \partial_\kappa E^\nu_m \partial_\mu \phi\right)
\right.\label{field_equation_E}\\&&\left.
-\frac{1}{2}\left(\partial_\kappa \partial_\mu \left(E \partial_\lambda E^{\mu m} E^\nu_m \partial_\nu \phi\right)
+\partial_\nu \left(E \partial_\lambda E^{\mu m} E^\nu_m \partial_\kappa \partial_\mu \phi\right)
+\partial_\kappa \partial_\nu \left(E \partial_\lambda E^{\mu m} E^\nu_m \partial_\mu \phi\right)
+\partial_\mu \left(E \partial_\lambda E^{\mu m} E^\nu_m \partial_\kappa \partial_\nu \phi\right)\right)
\right.\nonumber\\&&\left.
-\frac{1}{4} \partial_\lambda \partial_\mu \left(E E^{\mu m} E^\nu_m \partial_\kappa \partial_\nu \phi\right)
-\frac{1}{4} \partial_\kappa \partial_\nu \left(E E^{\mu m} E^\nu_m \partial_\lambda \partial_\mu \phi\right)
+\frac{1}{12}E \partial_\lambda E^\mu_a \partial_\kappa E^\nu_b R_{\mu\nu}^{ab}\left(E\right) \phi
+\frac{1}{6}E \partial_\lambda E^\mu_a E^\nu_b \partial_\kappa R_{\mu\nu}^{ab}\left(E\right) \phi
\right.\nonumber\\&&\left.
+\frac{1}{6}E \partial_\lambda E^\mu_a E^\nu_b R_{\mu\nu}^{ab}\left(E\right) \partial_\kappa \phi 
-\frac{1}{6}\partial_\kappa \left(E \partial_\lambda E^\mu_a E^\nu_b R_{\mu\nu}^{ab}\left(E\right)\phi\right)  
+\frac{1}{12}E E^\mu_a E^\nu_b \partial_\lambda R_{\mu\nu}^{ab}\left(E\right) \partial_\kappa \phi
-\frac{1}{12}\partial_\kappa \left(E E^\mu_a E^\nu_b \partial_\lambda R_{\mu\nu}^{ab}\left(E\right)\phi\right)
\right.\nonumber\\&&\left.
-\frac{1}{24}\partial_\lambda \left(E E^\mu_a E^\nu_b R_{\mu\nu}^{ab}\left(E\right) \partial_\kappa \phi\right)
-\frac{1}{24}\partial_\kappa \left(E E^\mu_a E^\nu_b R_{\mu\nu}^{ab}\left(E\right) \partial_\lambda \phi\right)
-\frac{1}{2\phi^2}E \partial_\lambda \phi
\partial_\kappa E^{\mu m} E^\nu_m \partial_\mu \phi \partial_\nu \phi
-\frac{1}{2}\partial_\lambda \left(\frac{1}{\phi}E
\partial_\kappa E^{\mu m} E^\nu_m \partial_\mu \phi \partial_\nu \phi\right)
\right.\nonumber\\&&\left.
-\frac{1}{2}\partial_\mu \left(\frac{1}{\phi}E \partial_\lambda \phi
\partial_\kappa E^{\mu m} E^\nu_m \partial_\nu \phi\right)
-\frac{1}{2}\partial_\nu \left(\frac{1}{\phi}E \partial_\lambda \phi
\partial_\kappa E^{\mu m} E^\nu_m \partial_\mu \phi \right)
-\frac{1}{12}\partial_\lambda\left(E \partial_\kappa E^\mu_a E^\nu_b R_{\mu\nu}^{ab}\left(E\right)\phi\right)
+\frac{1}{12}E \partial_\lambda \phi \partial_\kappa E^\mu_a E^\nu_b R_{\mu\nu}^{ab}\left(E\right)\right]
\nonumber\\&&
+\frac{1}{6}E E^\mu_a E^\nu_b \mathcal{R}_{\mu\nu}^{ab}\left(E,\phi,\theta\right)\phi
+\frac{1}{12}E E^\mu_a E^\nu_b \frac{\delta \mathcal{R}_{\mu\nu}^{ab}\left(E,\phi,\theta\right)}{\delta \phi}\phi^2
+2E C_{\mu\nu\rho\sigma}\left(E\right)\frac{\delta \mathcal{C}^{\mu\nu\rho\sigma}
\left(E,\phi,\theta\right)}{\delta \phi}
+\frac{\delta \mathcal{E}\left(E,\phi,\theta\right)}{\delta \phi}
\left[\frac{1}{2}E^{\mu m} E^\nu_m \partial_\mu \phi \partial_\nu \phi
\right.\nonumber\\&&\left.
+\frac{1}{12} E^\mu_a E^\nu_b R_{\mu\nu}^{ab}\left(E\right)\phi^2
+C_{\mu\nu\rho\sigma}\left(E\right)C^{\mu\nu\rho\sigma}\left(E\right)\right]
-\partial_\mu \left(\mathcal{E}\left(E,\phi,\theta\right) E^{\mu m} E^\nu_m \partial_\nu \phi\right)
+\frac{1}{6}\mathcal{E}\left(E,\phi,\theta\right) E^\mu_a E^\nu_b R_{\mu\nu}^{ab}\left(E\right)\phi
+\mathcal{O}\left(\theta^2\right)=0,\nonumber
\end{eqnarray}

\begin{eqnarray}
&&-E E_\mu^a\left\{\frac{1}{2} E^{\mu m} E^\nu_m \partial_\mu \phi \partial_\nu \phi
+\frac{1}{12} E^\mu_a E^\nu_b R_{\mu\nu}^{ab}\left(E\right)\phi^2
+C_{\mu\nu\rho\sigma}\left(E\right)C^{\mu\nu\rho\sigma}\left(E\right)
\right.\nonumber\\&&\left.
+i\theta^{\lambda\kappa}\left[
\frac{1}{4} \partial_\lambda E^{\mu m} \partial_\kappa E^\nu_m \partial_\mu \phi \partial_\nu \phi
+\frac{1}{2}\partial_\lambda E^{\mu m} E^\nu_m \partial_\kappa \partial_\mu \phi \partial_\nu \phi
+\frac{1}{2}\partial_\lambda E^{\mu m} E^\nu_m \partial_\mu \phi \partial_\kappa \partial_\nu \phi
+\frac{1}{4} E^{\mu m} E^\nu_m \partial_\lambda \partial_\mu \phi \partial_\kappa \partial_\nu \phi
\right.\right.\nonumber\\&&\left.\left.
+\frac{1}{24} \partial_\lambda E^\mu_a \partial_\kappa E^\nu_b R_{\mu\nu}^{ab}\left(E\right) \phi^2
+\frac{1}{12} \partial_\lambda E^\mu_a E^\nu_b \partial_\kappa R_{\mu\nu}^{ab}\left(E\right) \phi^2
+\frac{1}{6} \partial_\lambda E^\mu_a E^\nu_b R_{\mu\nu}^{ab}\left(E\right) \phi \partial_\kappa \phi 
+\frac{1}{12} E^\mu_a E^\nu_b \partial_\lambda R_{\mu\nu}^{ab}\left(E\right) \phi \partial_\kappa \phi
\right.\right.\nonumber\\&&\left.\left.
+\frac{1}{24} E^\mu_a E^\nu_b R_{\mu\nu}^{ab}\left(E\right) \partial_\lambda \phi \partial_\kappa \phi
+\frac{1}{2\phi}\partial_\lambda \phi
\partial_\kappa E^{\mu m} E^\nu_m \partial_\mu \phi \partial_\nu \phi
+\frac{1}{12}\partial_\lambda \phi
\partial_\kappa E^\mu_a E^\nu_b R_{\mu\nu}^{ab}\left(E\right)\phi \right]
+\frac{1}{12} E^\mu_a E^\nu_b \mathcal{R}_{\mu\nu}^{ab}\left(E,\phi,\theta\right)\phi^2
\right.\nonumber\\&&\left.
+2C_{\mu\nu\rho\sigma}\left(E\right)\mathcal{C}^{\mu\nu\rho\sigma}\left(E,\phi,\theta\right)\right\}
+\frac{\delta \mathcal{E}\left(E,\phi,\theta\right)}{\delta E^\mu_a}
\left[\frac{1}{2} E^{\mu m} E^\nu_m \partial_\mu \phi \partial_\nu \phi
+\frac{1}{12} E^\mu_a E^\nu_b R_{\mu\nu}^{ab}\left(E\right)\phi^2
+C_{\mu\nu\rho\sigma}\left(E\right)C^{\mu\nu\rho\sigma}\left(E\right)\right]
\nonumber\\&&
+E E^{\nu a} \partial_\mu \phi \partial_\nu \phi
+\frac{1}{6}E E^\nu_b R_{\mu\nu}^{ab}\left(E\right)\phi^2
+2E \frac{\delta C_{\lambda\nu\rho\sigma}\left(E\right)}{\delta E^\mu_a}C^{\lambda\nu\rho\sigma}\left(E\right)
+i\theta^{\lambda\kappa}\left[
-\frac{1}{4}\partial_\lambda \left(E \partial_\kappa E^{\nu a} \partial_\mu \phi \partial_\nu \phi\right)
-\frac{1}{4}\partial_\kappa \left(E \partial_\lambda E^{\nu a} \partial_\mu \phi \partial_\nu \phi\right)
\right.\nonumber\\&&\left.
+\frac{1}{2}E \partial_\lambda E^{\nu a} \partial_\kappa \partial_\mu \phi \partial_\nu \phi
-\frac{1}{2}\partial_\lambda \left(E E^{\nu a} \partial_\kappa \partial_\mu \phi \partial_\nu \phi\right)
+\frac{1}{2}E \partial_\lambda E^{\nu a} \partial_\mu \phi \partial_\kappa \partial_\nu \phi
-\frac{1}{2}\partial_\lambda \left(E E^{\nu a} \partial_\mu \phi \partial_\kappa \partial_\nu \phi\right)
\right.\nonumber\\&&\left.
+\frac{1}{4}E E^{\nu a} \partial_\lambda \partial_\mu \phi \partial_\kappa \partial_\nu \phi
+\frac{1}{4}E E^{\nu a} \partial_\lambda \partial_\nu \phi \partial_\kappa \partial_\mu \phi
-\frac{1}{24}E^\mu_a \partial_\lambda\left(E \partial_\kappa E^\nu_b R_{\mu\nu}^{ab}\left(E\right) \phi^2\right)
-\frac{1}{24}E^\mu_a \partial_\kappa \left(E \partial_\lambda E^\nu_b R_{\mu\nu}^{ab}\left(E\right) \phi^2\right)
\right.\nonumber\\&&\left.
+\frac{1}{24}E \partial_\lambda E^\rho_c \partial_\kappa E^\nu_b
\frac{\delta R_{\rho\nu}^{cb}}{\delta E^\mu_a}\left(E\right) \phi^2
+\frac{1}{12}\left(E \partial_\lambda E^\nu_b \partial_\kappa R_{\mu\nu}^{ab}\left(E\right) \phi^2
-\partial_\lambda \left(E E^\nu_b \partial_\kappa R_{\mu\nu}^{ab}\left(E\right) \phi^2\right)
-\frac{\delta R_{\rho\nu}^{cb}}{\delta E^\mu_a}\partial_\kappa\left(E \partial_\lambda E^\rho_c E^\nu_b
\left(E\right) \phi^2\right)\right)
\right.\nonumber\\&&\left.
+\frac{1}{6}\left(E \partial_\lambda E^\nu_b R_{\mu\nu}^{ab}\left(E\right) \phi \partial_\kappa \phi
-\partial_\lambda\left( E E^\nu_b R_{\mu\nu}^{ab}\left(E\right) \phi \partial_\kappa \phi\right)
+E \partial_\lambda E^\rho_c E^\nu_b \frac{\delta R_{\rho\nu}^{cb}\left(E\right)}{\delta E^\mu_a}
\phi \partial_\kappa \phi
+E E^\nu_b \partial_\lambda R_{\mu\nu}^{ab}\left(E\right) \phi \partial_\kappa \phi\right)
\right.\nonumber\\&&\left.
-\frac{1}{12} \frac{\delta R_{\rho\nu}^{cb}\left(E\right)}{\delta E^\mu_a}\partial_\lambda
\left(E E^\rho_c E^\nu_b  \phi \partial_\kappa \phi\right)
+\frac{1}{12}E E^\nu_b R_{\mu\nu}^{ab}\left(E\right) \partial_\lambda \phi \partial_\kappa \phi
+\frac{1}{24}E E^\mu_a E^\nu_b \frac{\delta R_{\mu\nu}^{ab}\left(E\right)}{\delta E^\mu_a}
\partial_\lambda \phi \partial_\kappa \phi
+\frac{1}{2\phi}E \partial_\lambda \phi
\partial_\kappa E^{\nu a} \partial_\mu \phi \partial_\nu \phi
\right.\nonumber\\&&\left.
-\partial_\kappa \left(\frac{1}{2\phi}E \partial_\lambda \phi
E^{\nu a} \partial_\mu \phi \partial_\nu \phi\right)
+\frac{1}{12}\left(E \partial_\lambda \phi
\partial_\kappa E^\nu_b R_{\mu\nu}^{ab}\left(E\right)\phi
-\partial_\kappa \left(E \partial_\lambda \phi
E^\nu_b R_{\mu\nu}^{ab}\left(E\right)\phi\right)
+E \partial_\lambda \phi \partial_\kappa E^\rho_c E^\nu_b \frac{\delta R_{\rho\nu}^{cb}\left(E\right)}
{\delta E^\mu_a}\phi\right)\right]
\nonumber\\&&
+\frac{1}{6}E E^\nu_b \mathcal{R}_{\mu\nu}^{ab}\left(E,\phi,\theta\right)\phi^2
+\frac{1}{12}E E^\rho_c E^\nu_b \frac{\delta \mathcal{R}_{\rho\nu}^{cb}\left(E,\phi,\theta\right)}{\delta E^\mu_a}\phi^2
+2E \frac{\delta C_{\lambda\nu\rho\sigma}\left(E\right)}{\delta E^\mu_a}
\mathcal{C}^{\lambda\nu\rho\sigma}\left(E,\phi,\theta\right)
+2E C_{\lambda\nu\rho\sigma}\left(E\right)\frac{\delta \mathcal{C}^{\lambda\nu\rho\sigma}\left(E,\phi,\theta\right)}
{\delta E^\mu_a}
\nonumber\\&&
+\mathcal{E}\left(E,\phi,\theta\right)\left[E^{\nu a} \partial_\mu \phi \partial_\nu \phi
+\frac{1}{6} E^\nu_b R_{\mu\nu}^{ab}\left(E\right)\phi^2
+\frac{1}{12} E^\rho_c E^\nu_b \frac{\delta R_{\rho\nu}^{cb}\left(E\right)}{\delta E^\mu_a}\phi^2
+2\frac{\delta C_{\lambda\nu\rho\sigma}\left(E\right)}{\delta E^\mu_a}
C^{\lambda\nu\rho\sigma}\left(E\right)\right]
+\mathcal{O}\left(\theta^2\right)=0.
\label{field_equation_phi}
\end{eqnarray}
In the field equations ($\ref{field_equation_E}$) and ($\ref{field_equation_phi}$),
the variation of the Riemann tensor, of the Weyl tensor, of their extenstions and of the extension
of the determinant of the terad field lead to boundary terms vanishing through integration.

\section{Summary and Discussion}

It has been formulated conformal gravity on noncommutative spacetime. The presupposed gravity action consists of the
conformal Weyl gravity action and a special Brans-Dicke action, where the prefactor of the coupling term between the
Ricci scalar and the scalar field has been chosen in such a way that its transformation cancels the transformation
of the kinetic term of the scalar field under conformal transformations. The geometry of the noncommutative spacetime
has been assumed to be described by canonical commutation relations of the coordinates. To treat the gravity action
as field theory on noncommutative spacetime, the moyal star product approach has been used, which is based on Weyl
quantization and maps products of fields depending on noncommuting coordinates to generalized products of these
fields depending on usual commuting coordinates and the tensor determining the nontrivial commutation relations
of the coordinates. If the conformal gravity action were reformulated by just transforming usual products to moyal
star products, conformal invariance would be spoiled, because the products within the transformation rules containing
the gauge parameter have to be transformed to star products, too. To maintain a certain local symmetry on noncommutative spacetime, conformal symmetry in the special case considered in this paper, Seiberg-Witten maps have to be introduced
mapping the fields to generalized fields depending on the original fields and on the noncommutativity tensor
and mapping the gauge parameter to a generalized gauge parameter.
Accordingly, the corresponding Seiberg-Witten maps have been formulated by using the consistency condition that a usual
gauge transformation of the generalized fields, what means a commutative gauge transformation of the usual fields
appearing in the expression of the Seiberg-Witten map, has to be equal to a noncommutative gauge transformation of
the generalized fields with the generalized gauge parameter. Whereas the Seiberg-Witten map of the scalar field maps
the usual scalar field to itself implying that the noncommuting quantity equals the usual quantity, the gauge parameter
as well as the tetrad field describing the gravitational field and the inverse tetrad field are modified decisively.
Replacing the usual coordinates by the noncommuting coordinates, replacing the corresponding products of fields by the
moyal star product and inserting the obtained Seiberg-Witten maps to the gravity action have yielded the generalized
gravity action from which the field equations of gravity can be extracted.   
Since the Seiberg-Witten map of the tetrad field also contains the scalar field, this leads to a very complicated
interaction structure between the tetrad field and the scalar field, because the generalized Ricci scalar as well
as the generalized Weyl tensor depend on the scalar field.
As theory on noncommutative spacetime the presented conformal gravity theory could be seen as a classical
approximation to a fundamental quantum theory of gravity, where the problem of nonrenormalizability does perhaps
not appear. If the conformal symmetry would be interpreted as fundamental symmetry, the theory should be combined
with a conformal invariant version of the standard model. In combination with a further scalar field, the Higgs
field of the standard model coupled in an appropriate way to the other scalar field for example, and under
presupposition of self-interaction terms of fourth order for the scalar fields, the gravitational constant
could be generated by spontaneous symmetry breaking without violating conformal invariance and this would
lead to the usual Einstein-Hilbert term on noncommutative spacetime. The application of the modified gravity
theory to cosmology could yield a possible explanation of the acceleration of the expansion rate of the universe.


\end{document}